\documentclass[a4paper]{article}
\usepackage[a4paper,vmargin={30mm,30mm},hmargin={25mm,25mm}]{geometry}
\usepackage{setspace}
\usepackage{graphicx}
\usepackage{epstopdf}

\begin{document}

\onehalfspacing
\title{A high-power waveguide resonator second harmonic device with external conversion efficiency up to 75\,\%.}
\author{M. Stefszky, R. Ricken, C. Eigner, V. Quiring, H. Herrmann and C. Silberhorn}

\maketitle

\begin{abstract}
We report on a highly efficient waveguide resonator device for the production of 775\,nm light using a titanium indiffused LiNbO$_3$ waveguide resonator. When scanning the resonance the device produces up to 110\,mW of second harmonic power with 140\,mW \textit{incident} on the device - an external conversion efficiency of 75\%. The cavity length is also locked, using a Pound-Drever-Hall type locking scheme, involving feedback to either the cavity temperature or the laser frequency. With laser frequency feedback a stable output power of approximately 28\,mW from a 52\,mW pump is seen over one hour.
\end{abstract}

\section{Introduction}

Integrated optical waveguide technologies offer improvements in size, scalability, integrability and nonlinear efficiency over their bulk counterparts. These properties have been used to create a myriad of impressive devices, such as high-speed switches \cite{Wooten00.JQE}, optical parametric oscillators \cite{Arbore97.OL}, sources of optical squeezing \cite{Anderson95.OL,Stefszky17.PRA} and frequency conversion devices \cite{Ruetz16.APB}. Frequency conversion is often used to produce laser fields at wavelengths inaccessible to current materials \cite{Iwai03.APL}, but can also be used to interface individual parts of a network \cite{Ruetz16.APB} or to transfer information from one field to another at a different wavelength \cite{Vandevender.JOMO}.

Traditionally, the main advantage of waveguide devices has been very high nonlinear efficiency, allowing for relatively high conversion efficiencies at low input and low output powers. However, for many applications both high conversion efficiency and high output powers are desired. For example recent work has shown that one can use high performance second harmonic (SH) generation to up-convert the frequency of a squeezed vacuum field \cite{Vollmer14.PRL}. Furthermore, observations of frequency comb generation in second-order free-space resonator devices \cite{Ulvila13.OL,Ricciardi15.PRA} has driven further understanding of these systems \cite{Leo16.PRL} and are generating demand for high performance, high power, second-order devices. The integration of devices such as these is the natural step towards moving these technologies into the commercial realm. Unfortunately, the performance of waveguides has traditionally been limited due to a number of factors such as  large waveguide losses, poor mode shape profile and high power effects such as photorefraction. Only recently have large improvements in stable, high conversion efficiency and high power waveguides been realised. The most common architecture used to this aim is MgO:LiNbO$_3$ ridge waveguides \cite{Sun12.OL,Iwai03.APL,Mizuuchi03.OL,Sakai07.OL,Sakai06.OL}. Sun \textit{et al} have recently produced single-pass waveguides with internal second harmonic conversion efficiencies of around 70\% and with output powers exceeding 400\,mW at 532\,nm \cite{Sun12.OL}.

One can further enhance the strength of the nonlinear interaction by resonating the optical field within the waveguide. This method of enhancement has been demonstrated by various groups in a number of different architectures, albeit at low operational powers \cite{Luo15.NJP,Regener88.JOSAB,Pernice12.APL,Scaccabarozzi06.OL,Fujimura96.JQE}.    Resonators also allow one to shape the spectrum of the output light \cite{Luo15.NJP} and to tailor the system for optimum conversion for a given pump power and waveguide loss \cite{Regener88.JOSAB}. Despite these benefits, investigation of waveguide resonators for \textit{high power} and \textit{high conversion efficiency} applications has not been experimentally realised. This is likely due to increased sensitivity to temperature fluctuations (over their single-pass counterparts) as well as traditionally high waveguide losses. Titanium indiffused waveguides exhibit the lowest losses observed from any nonlinear waveguide technology, with losses as low as 0.02\,dB/cm \cite{Luo15.NJP}. Ridge waveguides have shown measured losses as low as 0.1\,dB/cm, \cite{Gui09.OE}. At the same time, ridge waveguides have demonstrated single-pass second harmonic conversion efficiencies of around 240\%W$^{-1}$cm$^{-2}$ (for SH at 340nm) \cite{Mizuuchi03.OL} whilst titanium indiffused waveguides produced in our group have shown single-pass second harmonic conversion efficiencies up to around 40\%W$^{-1}$cm$^{-2}$ in short samples.

Due to the low losses exhibited in titanium indiffused waveguides, the resonator was produced using this technology. Indiffused waveguides also have a very high spatial overlap with single mode fibres (of up to around 94\%, necessary for achieving a high external conversion efficiency). The major disadvantage of titanium indiffused waveguides is the presence of photorefraction \cite{Nava13.APL,Pal15.APB,Becker85.APL,Kondo.AO}. Whilst this effect can be compensated for at lower operational powers, catastrophic photorefraction will occur at higher powers and will produce large amounts of noise on the produced SHG \cite{Carrascosa08.OE}.

In this paper we present the realisation of a high power waveguide resonator, whose external second harmonic conversion efficiency (defined here as the amount of second harmonic power exiting the device divided by the amount of fundamental power incident on the device) is shown to be as high as 75\%. The resonator length has also been locked using a Pound-Drever-Hall (PDH) type locking scheme \cite{Black01.AJP}, resulting in a reduced conversion efficiency of slightly more than 50\,\%, but with stable output power of up to around 30\,mW over an hour. 

The device highlights the previously unadvertised performance capabilities and limitations of titanium indiffused waveguides as resonators in applications where high power and low losses are important. Whilst the wavelength of the converted light form this particular device is easily accessible by various laser technologies, it does provide access to a phase congruent fundamental (which lies in the communications band) and second harmonic field. This is useful, for example, in nonlinear optics experiments that pump any process with the second harmonic of the field such as parametric amplification and oscillation \cite{Arbore97.OL}, as well as detection of these states using, for example, balanced homodyne detection \cite{Stefszky12.CQG,Mehmet11.OE}.

\section{The waveguide SHG device}

Waveguides are fabricated in z-cut LiNbO$_3$ by an indiffusion of lithographically patterned 7 $\mu$m wide, 80 nm thick titanium strips at 1060 $^\circ$C for 9 hours. The diffusion takes place in a platinum box in a diffusion tube that is flushed with oxygen in order to mitigate LiO$_2$ outdiffusion. An insulating photoresist pattern is then applied via a second lithography step for the field assisted periodic domain inversion. The sample is finally annealed at 300 degrees Celsius for 2 hours to reduce stress at the domain walls. We have chosen a poling period of approximately 16.9 $\mu$m  to achieve the desired Type 0 quasi-phase matching at a temperature of around 180 degrees Celsius, in order to reduce photorefractive damage \cite{Carrascosa93.JAP}. A short sample length of 8 mm long is chosen as a smaller sample is easier to stabilise in temperature and ensures that phase matching is uniform across the entire sample. For all measurements the sample is heated to the ideal phase matching temperature of around 180 degrees Celsius.

\begin{figure}[!ht]
\centering
  \includegraphics[width=0.8\linewidth]{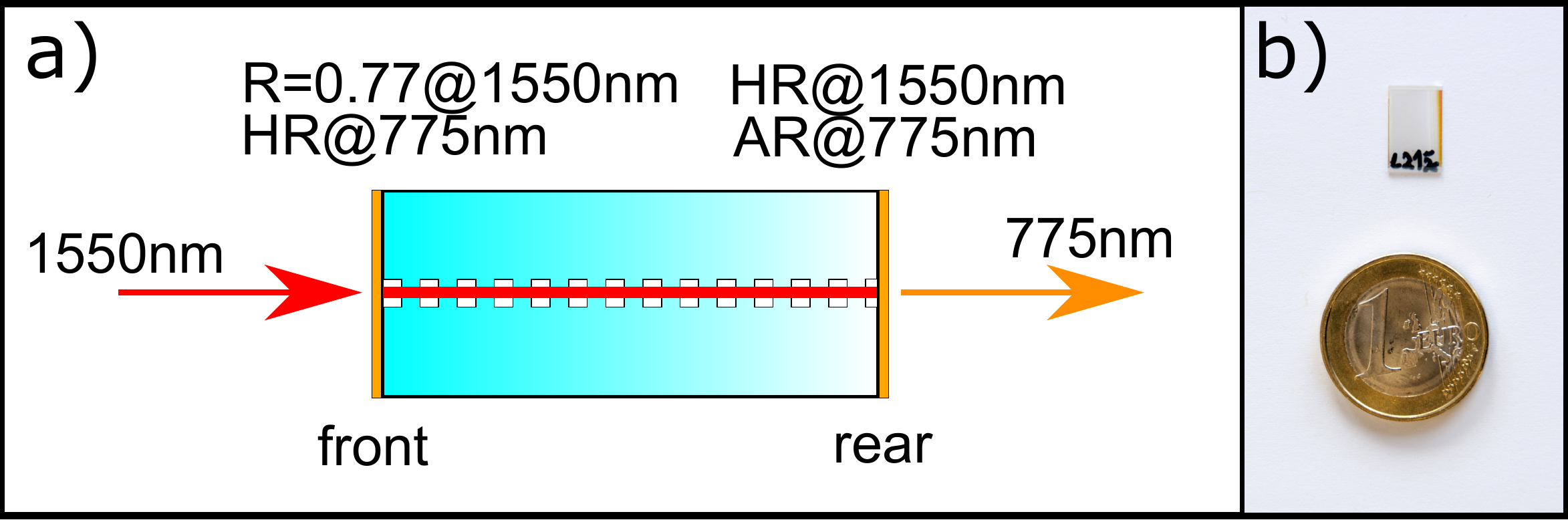}
  \caption{A schematic of the device is shown in a). A pump field at 1550nm is incident on the front coating of the device and the generated second harmonic escapes the sample from the coating on the rear. Only a single waveguide and a representation of its periodic poling is shown and is not to scale. A photograph of a similar waveguide sample is shown in b).}
  \label{sample}
\end{figure}



In order to optimise the conversion efficiency, one needs to correctly choose the mirror coating reflectivities. We choose to have the second harmonic exit the rear of the sample whilst the pump enters through the front. The rear surface therefore has an anti-reflection (AR) coating at 775\,nm and a high-reflectivity (HR) coating for 1550\,nm. The front surface is coated with a 77\% mirror at 1550\,nm and a HR coating at 775\,nm, thereby resulting in a double-pass configuration for the SH field \cite{Imeshev98.OL}. The 77\% coating is chosen to achieve critical coupling of the pump assuming intra-cavity losses of 0.07dB/cm and a pump power of approximately 50\,mW using the expected nonlinear properties of the device \cite{Regener88.JOSAB}. The coatings for the device are produced in-house using ion-beam assisted vapour deposition. Lithium niobate ``witness samples'' are coated in the same coating runs and from these one can accurately determine the properties of the coatings. The HR coatings are expected to have a reflectivity greater than 99$\%$, and AR coatings a reflectivity of around 0.5$\%$. We note that the performance of the double-pass SH device depends critically on the phase relationship between the forward and reverse-propagating fields. To address this issue, the waveguide sample consists of  many (around 100) waveguides. The polishing of the waveguide sample end-faces is not perfectly parallel and therefore the phase relationship varies from waveguide to waveguide and provides a means for searching over this parameter \cite{Imeshev98.OL}.

\section{Waveguide resonator characterisation}

\subsection{Linear properties}

We begin by determining the linear properties of the device at the fundamental wavelength of 1550nm. Using the standard Fabry-Perot equations one can find expressions for the ratios between the transmitted power and the incident power $(P_{trans})/(P_{in})$ as well as the reflected power and the incident power on resonance $(P_{ref}^{on})/(P_{in})$ \cite{Siegman},
\begin{eqnarray}
\frac{P_{ref}^{on}}{P_{in}} &=& \frac{(\sqrt{R_{f}}-\sqrt{R_{r}} e^{- \alpha L})^2}{(1-\sqrt{R_{f} R_{r}} e^{- \alpha L})^2} \label{ratioref}\\
\frac{P_{trans}}{P_{in}} &=& \frac{(1-R_{f})(1-R_{r}) e^{- \alpha L}}{(1-\sqrt{R_{f} R_{r}} e^{- \alpha L})^2}, \label{ratiotrans}
\end{eqnarray}
where $L$ is the sample length in centimetres, $\alpha$ is the loss rate per centimeter, $R_f$ is the power reflectivity for the front mirror and $R_r$ is the power reflectivity for the rear mirror. If the reflectivity of the front mirror is well known, then these equations can be solved to find the intra-cavity loss rate and the reflectivity of the HR mirror.

\begin{figure}
\centering
  \includegraphics[width=0.8\linewidth]{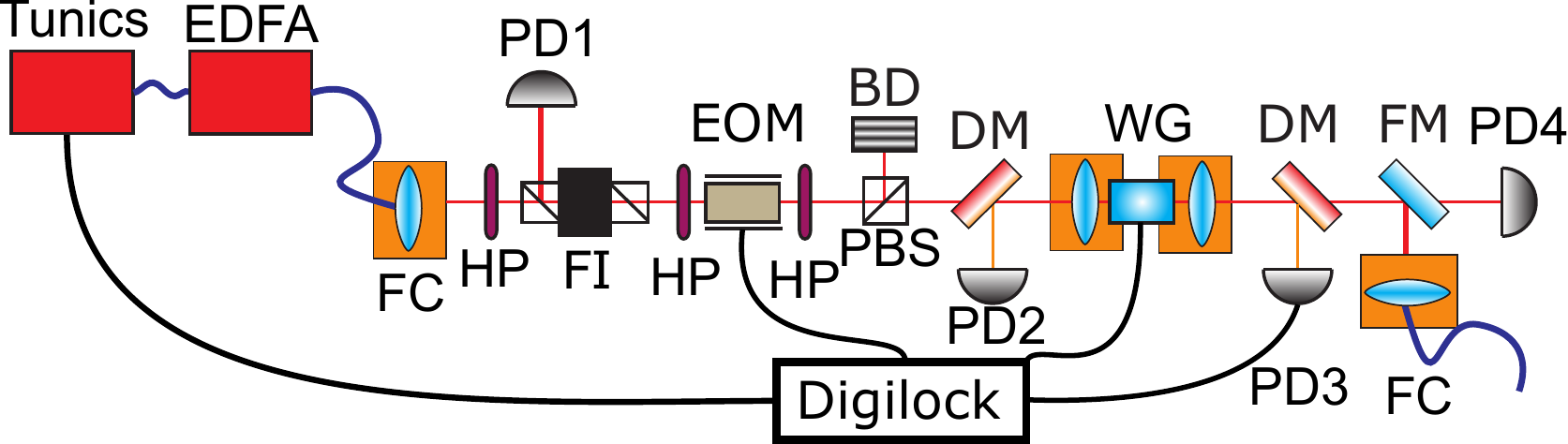}
  \caption{Experimental layout. Tunics: Tunics 1550nm tunable laser, EDFA: Erbium-doped fibre amplifier, FC: Fibre coupler, PD: Photodetector, FI: Faraday isolator, HP: Half-wave plate, EOM: Electro-optic phase modulator, PBS: Polarising beam-splitter, BD: Beam dump, WG: Temperature-controlled waveguide, DM: Dichroic mirror, FM: Flipper mirror.}
  \label{experiment}
\end{figure}

The experimental layout is shown in Fig. \ref{experiment}. The waveguide is pumped by a Tunics tunable C-band laser that is amplified using an EDFA. For this measurement the wavelength of the laser is offset by 10nm, such that the phase matching condition is \textit{not} met, thereby removing the nonlinear interaction. The light passes through a Faraday isolator that is used to measure the light reflected by the cavity, at PD1. The light then passes through an EOM, later used in combination with the Toptica digilock system (which consists of a high voltage amplifier, a mixer and a waveform generator)for locking the cavity length. The feedback signal from this locking system can be connected to the waveguide heater or the laser frequency control. The power can then be fine-tuned using a half-wave plate followed by a polarizing beam splitter. A dichroic mirror before the front surface of the waveguide can be used to measure the amount of second harmonic light leaking from the coating on this surface incident at PD2. A second dichroic mirror at the exit of the waveguide separates the transmitted pump light, which can be measured at PD4, from the generated SH light, which is measured at PD3. All of the relevant light powers are then measured using a Thorlabs PM100 power meter and are cross-referenced using the photodetectors in the setup, ensuring that all results are consistent.

For accurate conversion efficiency measurements the mode overlap between the (fibre-coupled) laser mode and the waveguide mode is first precisely determined. The coupling of the laser beam into the mode of the waveguide is optimised and the light transmitted through the cavity is then coupled into a fibre as shown in Fig. \ref{experiment}. The fiber containing the laser field is then disconnected from the experiment and reconnected to this rear fibre-coupling, resulting in the laser field entering the waveguide from the rear surface. The Faraday isolator is then removed such that the light transmitted through the front face of the cavity (and therefore in the cavity spatial mode) is incident upon the fibre coupler. One can then measure the amount of power incident on and exiting the first fibre-coupler to determine a lower bound for the mode overlap. From this measurement, the overlap between the fibre mode and the waveguide mode is determined to be 0.93$\pm 0.02$.

Next, approximately 300$\mu $W of power is coupled into the waveguide cavity. The laser frequency is scanned and on reflection we find that the power on resonance is 36$\pm 1 \%$ of the power off resonance. We also find that the transmitted power on resonance is 6.3$\pm 0.1 \%$ of the power entering the cavity. Inserting these values into the ratios derived from Eq \ref{ratioref} and Eq \ref{ratiotrans}, one finds that the intra-cavity loss at the fundamental wavelength is 0.16$\pm 0.01$ dB/cm and the reflectivity of the rear mirror at the fundamental wavelength is 99.4$\pm0.1 \%$. Using these values one can now determine the cavity finesse $\mathit{F} = \frac{\pi \sqrt{r}}{1-r}=20$, free spectral range $FSR = \frac{c}{2 L n} = 8.7$GHz and cavity linewidth $\nu = \frac{FSR}{\mathit{F}} = 440$\,MHz; where $L$ is the sample length, $n$ is the refractive index of the fundamental wave and $r = \sqrt{R_f R_r R_l}$. The loss is unfortunately higher than desired, this is due to the fact that it is necessary to find a waveguide with the right phase matching condition, a high nonlinear conversion efficiency and low losses. This waveguide is the best compromise of the three parameters on this particular device.

\subsection{Conversion efficiency}

\subsubsection{Non pump-depleted regime}
With the fundamental cavity well characterized, attention is now turned to the nonlinear properties of the device. We begin our treatment in the low-power regime, where analytic solutions can be found and from these the desired parameters can be determined. To further simplify the treatment, it is assumed that the coating on the front face does not transmit any power at 775$\,$nm and that the coating on the rear face perfectly transmits power at this wavelength.  The normalized (single-pass) conversion efficiency, $\eta_{norm}$, is first defined,
\begin{eqnarray}
\eta_{norm} &=& \frac{P_{SH}}{P_{in}^2 L^2},
\end{eqnarray}
where $P_{SH}$ is the SH power exiting the waveguide resonator at the rear surface, $P_{in}$ is the pump power entering the sample and $L$ is the sample length. 

The resonator provides an intra-cavity build-up of the pump field, which can be characterized with the build-up factor, $f$ \cite{Regener88.JOSAB},
\begin{eqnarray}
f &=& \frac{1-R_f}{(1-\sqrt{Rf Rr}e^{-\alpha L})^2},
\end{eqnarray}
where $\alpha$ is the waveguide loss at the fundamental. This build-up factor can be combined with the single-pass normalized conversion efficiency to determine the conversion efficiency of the resonant device. The conversion efficiency (in the absence of pump depletion) for the double-pass pump-resonant device described here is defined \cite{Imeshev98.OL,Regener88.JOSAB},
\begin{eqnarray}
\frac{P_{SH}}{P_{F}} &\approx & 4 \eta_{norm} h P_{in} f^2 (1-cos(\theta)),
\label{conveffeq}
\end{eqnarray}
where $\theta$ represents the relative phase shift between the SH and fundamental fields due to the HR mirror at 775nm on the front surface of the device and the width of the domain adjacent to this coating. The factor of 4 in Eq. \ref{conveffeq} represents the fact that perfect constructive interference of the forward and reverse fields results in a conversion efficiency that is identical to a device operated at the same power with a length of $2L$. The waveguide length is replaced by the loss-corrected form, $h$, and is given by \cite{Regener88.JOSAB},
\begin{eqnarray}
h &=& \left( \frac{ 2 e^{-\alpha L} - e^{-\alpha_{2 \omega}L /2}}{\alpha-\alpha_{2\omega}/2} \right)^2,
\end{eqnarray}
where $\alpha_{2\omega}$ is the waveguide loss for the SH wave. Note that $h$ reduces to $L^2$ in the limit that the losses go to zero.

The conversion efficiency of the device is determined by measuring the SH power exiting the rear surface of the device as the pump power is varied. The pump power entering the waveguide is varied through rotation of a half-waveplate in the beam path and through varying the fibre amplifier gain. The frequency of the laser is scanned (using a triangle function with an amplitude that corresponds to a laser frequency shift of 3\,GHz at a scanning rate of approximately 20\,Hz) such that the laser frequency passes through one resonance of the cavity. The central wavelength of the laser is then shifted such that the optimal cavity resonance, the one that is closest to the centre of the phase-matching curve, is found. The central wavelength of the laser is then fine-adjusted at each power setting to ensure that thermal effects do not shift the cavity from the phase-matching condition. The generated second harmonic power on resonance is detected on PD3 after attenuation via a calibrated neutral density filter. These results are cross-checked for consistency using power meter measurements.

\begin{figure}
\centering
  \includegraphics[width=0.8\linewidth]{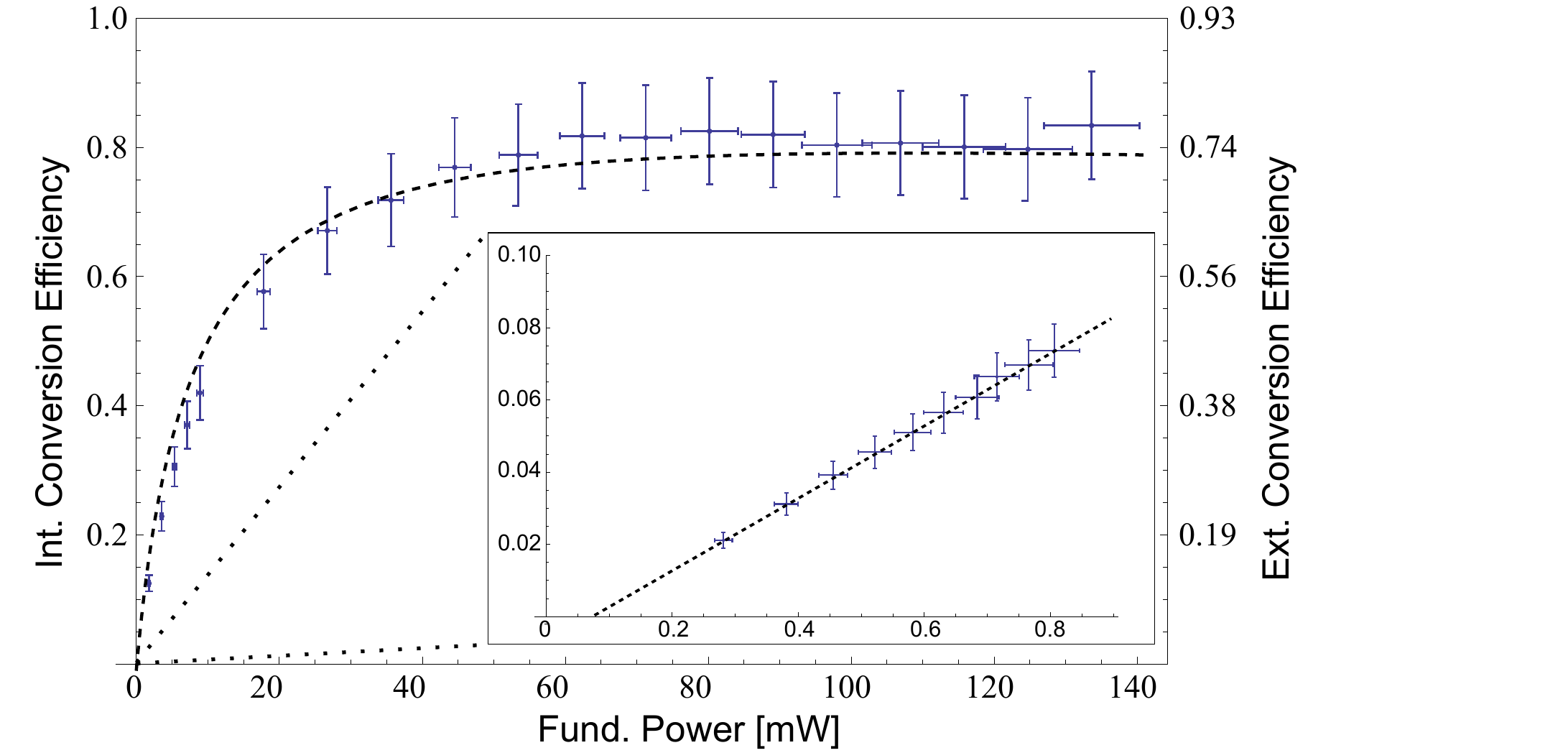}
  \caption{Measured conversion efficiency of the waveguide resonator. Data points are shown with error bars. The dashed line shows the theoretical fit using the losses measured in the previous section, assuming perfect constructive interference of the forward and reverse-propagating waves, and a normalized conversion efficiency of $38\%W^{-1}cm^{-2}$. Waveguide loss at the SH is assumed to be equal twice that found for the fundamental (of 0.16\,dB/cm). }
  \label{conveff}
\end{figure}

Fig. \ref{conveff} shows the measured second harmonic power levels as the input pump power is varied. A high power run is shown in the main figure, while an inset shows a zoom-in on the low-power regime. A fit to the data using the model described in Eq. \ref{conveffeq} is added to the inset, where the internal conversion efficiency is defined as the second harmonic power exiting the waveguide divided by the pump power matching the waveguide mode. Note that the relative phase shift $\theta$ and exact magnitude of the normalised conversion efficiency both affect the conversion efficiency of the device and cannot be independently determined. However, if we assume that the phase shift gives complete constructive interference $\theta = 90$ degrees, then the best fit of the conversion efficiency (using Eq. \ref{conveffeq}) requires a normalised nonlinear conversion efficiency of $38\%/(W.cm^2)$. This value coincides with the expected value from both theory \cite{Regener88.JOSAB} and previous measurements of other samples with similar length, thereby giving us confidence in the initial assumption about the phase. Using mode size measurements from similar waveguides and RSoft simulations, which give us an expected effective interaction area \cite{Regener88.JOSAB} of 85 $\mu$m$^2$, we determine the nonlinear coefficient to be $d_{eff} = 26$ pm/V. The $d_{eff}$ is not precisely known at 1550 nm but it has been measured to be 20 pm/V at 1330nm and up to 34 pm/V at shorter wavelengths\cite{Shoji97.JOSAB}.

\subsubsection{Pump depletion regime}

The expected performance of the device at higher powers (the main part of Fig. \ref{conveff}), where the non-pump depleted approximation is no longer valid, is found by numerically solving the coupled-mode equations in the same way as presented by Fujimura {\it et al} \cite{Fujimura96.JLT}. The coupled-mode equations for perfect phase matching are written,

\begin{eqnarray}
\frac{d}{dz}A(z) = -i \kappa^* B(z)A(z)^*-\frac{\alpha}{2}A(z)\\
\frac{d}{dz}B(z) = -i \kappa [A(z)]^2-\frac{\alpha_{2\omega}}{2} B(z),
\end{eqnarray}
where $A(z)$ and $B(z)$ are the complex amplitudes for the fundamental and second harmonic waves, respectively, and $\kappa$ is the nonlinear coupling coefficient. We assume here that $\kappa$ is real because the low-power modeling has shown that our system is well modeled under the assumption of perfect constructive interference between the forward and reverse traveling waves, or in other words, no phase shift on the quasi-phase matching grating.

A brief summary of the iterative method is described here, a full description can be found in \cite{Fujimura96.JLT}. At each end-face the internal fields are separated into forward and reverse traveling components. By assuming some initial conditions the coupled equations can be solved. These solutions are then checked for self-consistency with the initial conditions to ensure a meaningful result. If the fields are not self-consistent then a new initial condition is calculated based on the result of the previous iteration until self-consistency is reached. After many iterations a self-consistent solution for the internal fields is found, from which the external fields can be calculated using the boundary conditions. Using this method the conversion efficiency of the device was calculated and is shown in the main part of Fig. \ref{conveff}. The nonlinear coupling coefficient $\kappa = 0.71$ was found from a fit to the low-power measurements and all other parameters are the same as used in the low-power modeling. Note that the high power theory curve therefore has no free parameters, and still reasonably predicts the behavior of the system.


The external conversion efficiency is defined as the second harmonic power exiting the waveguide divided by the input pump power directly before the waveguide in-coupling lens. Note that the pump power measured in this way includes losses due to imperfect overlap between the pump and waveguide modes. It can be seen that an external conversion efficiency of 75\% is achieved for input powers of around 70\,mW, which we stress is achieved when the frequency of the laser is scanned over the cavity resonance. In Section \ref{Stability} it will be shown that we are unable to stabilise the cavity length at this power, nevertheless this scanning mode of operation may be useful for some applications. At fundamental input powers of 140\,mW and above the maximum transmitted power level seen as the laser was scanned over resonance was observed to rapidly fluctuate. This is perhaps an indicator of the onset of catastrophic photorefraction, but may also be due to other instabilities arising due to the high power levels, such as third-order nonlinearities.

\section{High power limitations}

The high power performance of the device is limited by photothermal and photorefractive effects. These effects are difficult to distinguish due to the fact that they are both expected to shift the cavity resonance in a similar fashion \cite{Fujiwara89.APL}. It is expected that as the device produces more power the magnitude of these effects will increase and the task of keeping the resonator on resonance will become more difficult.

Whilst it is difficult to get a quantitative analysis of these effects, it is possible to look at them qualitatively. By scanning the frequency of the laser one can scan over the cavity resonance approaching from both high and low frequency sides while detecting the transmitted power on PD4 (see Fig. \ref{experiment}). The presence of photothermal heating and/or photorefraction will produce a cavity scan that is asymmetric. This is because as resonance is approached from one side, heating of the cavity will tend to bring the cavity closer to resonance, leading to a very rapid transition to resonance. Approaching resonance from the opposite side will then cause heating that pushes the system away from resonance.

\begin{figure}
\centering
  \includegraphics[width=0.7\linewidth]{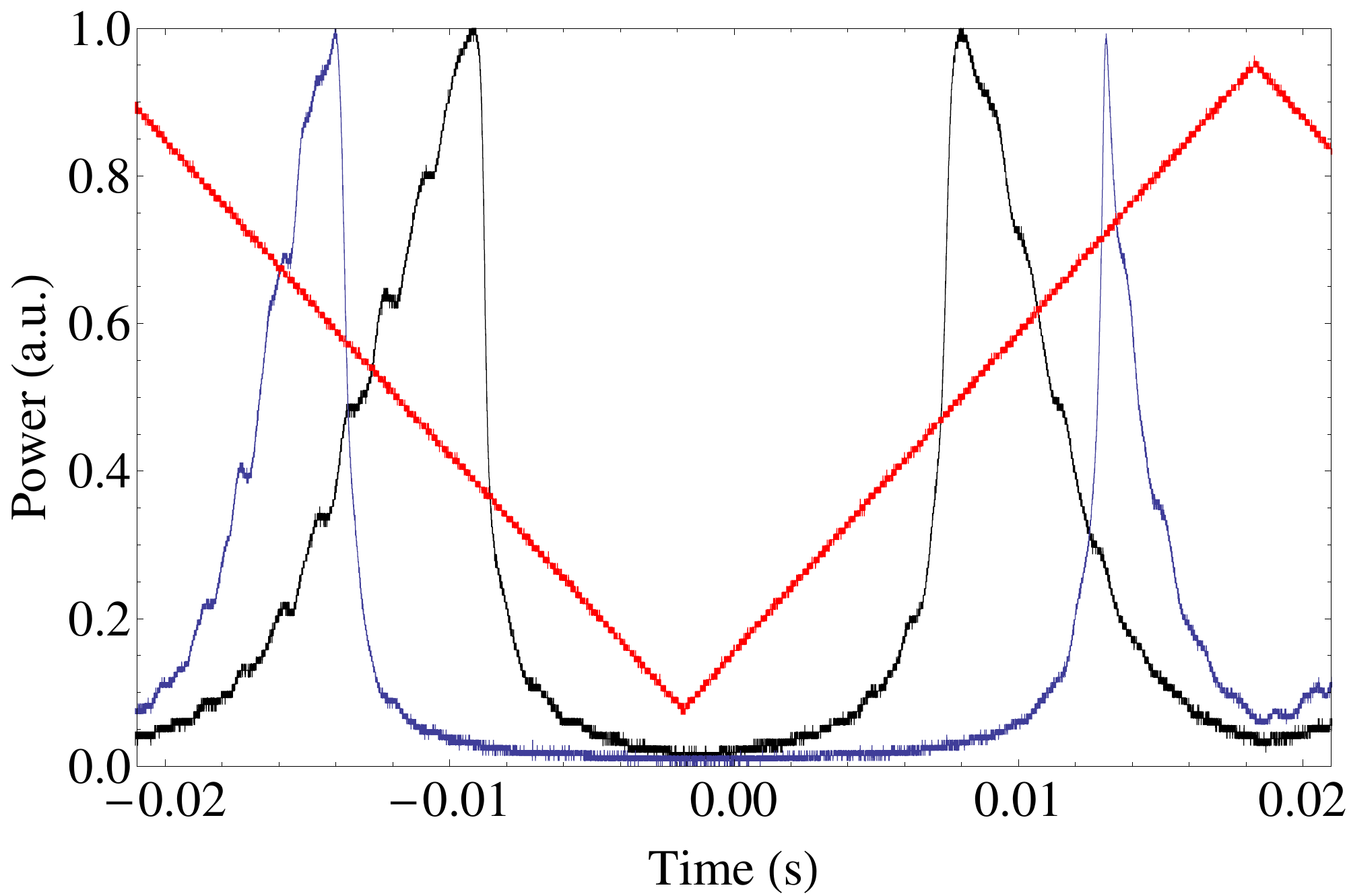}
  \caption{Cavity scan with approximately 110\,mW of fundamental power entering the sample. The blue trace is with the temperature far from phase matching while the black trace is on phase matching. The frequency of the laser is scanned with a triangle function, shown in red, the magnitude of which corresponds to a laser frequency shift of 3\,GHz. The oscillations on top of the cavity scan are due to interferences caused by the frequency scan.}
  \label{cavscan}
\end{figure}

Such a cavity scan is shown in Fig. \ref{cavscan}, where approximately 110\,mW of fundamental light is incident on the cavity. The direction of the laser frequency scan (scanned using a triangle function) is reversed at around zero seconds. The black trace shows the transmitted fundamental power when the cavity temperature is held at phase matching, resulting in a second harmonic power of around 70\,mW. We note that the expected Airy function describing the transmitted power through the cavity is not seen. Instead an asymmetric function of transmitted power is found, indicating some minor amount of photothermal and/or photorefractive effects.

To investigate the wavelength dependence of the asymmetry the frequency of the laser was varied, such that the device was far from phase matching and no second harmonic power was produced. The transmitted power was again measured as the cavity was scanned and is shown by the blue trace. One immediately notices that the cavity bandwidth has reduced due to the absence of nonlinear interaction, which reduces the fundamental field cavity decay rate, as expected. Perhaps surprisingly, however, a large asymmetry is still seen. Photorefraction is typically expected to have a very strong dependence on wavelength, whilst one might expect absorption to have a slight dependence on wavelength over this range. Therefore the observed results are consistent with a similar absorption at both wavelengths and only a minor contribution from photorefraction effects on the timescales of the scan. This is a surprising result and although it is outside the scope of this paper, it warrants further investigation.

\section{Stability}
\label{Stability}

The stability of frequency conversion devices is a critical factor in many applications. One typically requires device stability that is at least equal to the required measurement time. Therefore, a preliminary investigation into device stability over longer time periods is undertaken. Note that there is very little prior information regarding the long-term stability of high-power waveguide resonator.  Due to the narrow linewidth of the cavity resonance this requires very stable, fine temperature tuning and for greater stability, some form of feedback that locks to the cavity resonance.

The issue of stability is addressed in two steps. The first step is to optimise the standard temperature loop that heats the sample to around 180 degrees Celsius. The 2-stage oven is enclosed in a Teflon housing and is then encased in a larger housing with openings only for the input and output fields. With this setup the temperature of the cavity is stable to a couple of millikelvin or better and this is enough to ensure that the cavity resonance is held, at low input powers and without the PDH loop, on the minute time scale. This scheme is limited because it does not sense the resonance condition. As such, system drifts may force the cavity off resonance in a way that is not sensed by the temperature sensor in this loop and are therefore not compensated. A second step is required to provide stability over longer time scales and at higher powers.

The second step is the PDH locking loop which directly senses the resonance condition (of the fundamental field), thereby providing a means of compensating system drift over longer times. This locking loop was first achieved for low powers. The field entering the waveguide resonator is phase modulated via an electro-optic modulator at 25 MHz, as shown in Fig. \ref{experiment}. These sidebands are well within the cavity linewidth and as such are transmitted through the cavity with the carrier and are detected at PD3. This signal is then fed into a Toptica Digilock system, therein undergoing mixing and filtering to produce the error signal. This error signal is then used to lock the system to resonance via feedback. It is possible to achieve this feedback via two methods; the first is by feeding back to the temperature controller of the waveguide, and the second by feeding back to the laser frequency through actuation of a piezoelectric transducer. Feeding back to the laser provides a much faster lock but results in the centre frequency of the laser shifting, which will be an unwanted effect for some applications.

\begin{figure}[!ht]
\centering
  \includegraphics[width=0.7\linewidth]{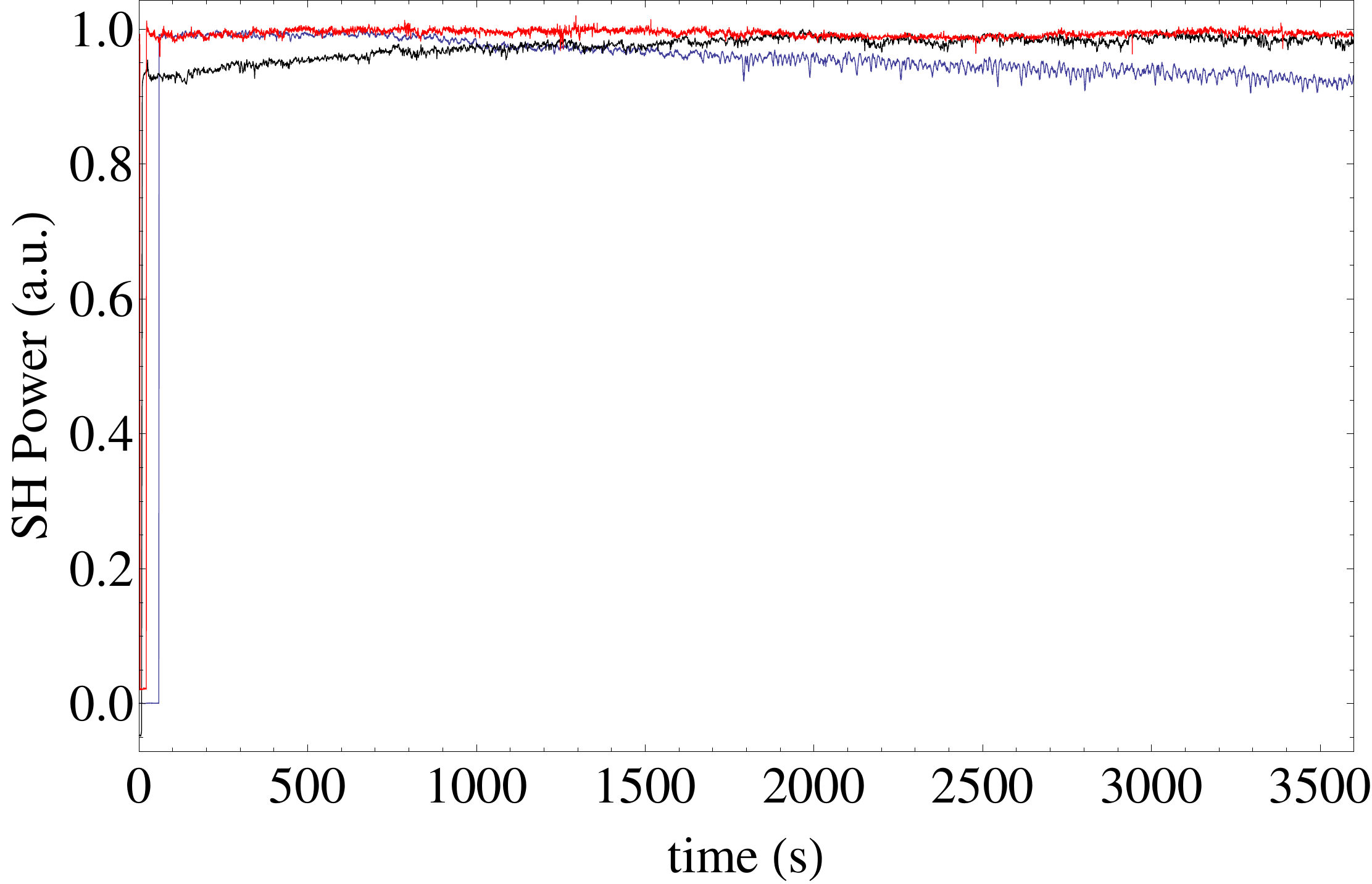}
  \caption{Output power of the SHG device over one hour with a temperature (blue) feedback loop and a low-power (red) and high-power (black) laser feedback loop enabled. The output SH power is normalised to the amount of power seen after the system has stabilised over a few minutes.}
  \label{lock}
\end{figure}

The performance of the two low-power locking schemes is illustrated in Fig. \ref{lock}. The temperature feedback lock (blue trace) is performed with approximately 30$\,$mW of fundamental field entering the cavity, resulting in approximately 10$\,$mW of second harmonic power generation, whilst the laser feedback scheme (red trace) has approximately $22\,$mW of fundamental field entering the cavity and 9$\,$mW second harmonic power exiting the system. As expected, the temperature feedback scheme does not perform as well as the laser frequency feedback as an accumulated offset is observed, likely due to the cavity resonance condition shifting away from the peak of the phase matching as the localized temperature of the waveguide shifts due to heating of the sample. In contrast, at the powers used here the laser feedback scheme locks to the maximum power immediately and has been locked stably for more than 6 hours, the first hour of which is illustrated in Fig. \ref{lock}.

In order to achieve a more stable lock at higher pump powers, the quality of the lock has to be improved. The Tunics laser is replaced with a New Focus velocity laser, providing a greater tuning range for the laser feedback. Furthermore, an integrated phase modulator allows for higher frequency modulation (of 302$\,$MHz), improving the capture range and allowing for detection of these sidebands on reflection of the cavity where the beat signal is stronger (with a fiber integrated photodiode replacing PD2). Approximately 52$\,$mW of fundamental field enters the cavity and the SH power stabilizes to around 28$\,$mW of SH after approximately 20 minutes, as illustrated by the black trace of Fig. \ref{lock}. This level of combined performance; high conversion efficiency, high power, and long-term stability, has not been presented previously from a waveguide resonator and highlights the versatility and quality of this device.

It is not currently possible to stably lock the SH cavity at higher input pump power. The output SH power of the locked cavity is seen to oscillate between two output SH modes at pump powers higher than around 50$\,$mW. This effect requires a drastic change in the phase-matching properties of the system and therefore may be due to localised photorefraction or even a Kerr effect. Operation at higher powers will require an understanding of this effect if one wishes to operate the system with a stable continuous-wave output.

\section{Conclusion}

In conclusion, we demonstrated a high external conversion efficiency, high power, second harmonic waveguide resonator device. This was achieved in an 8\,mm long titanium indiffused lithium niobate waveguide device. With a 1550\,nm pump field, an external conversion efficiency of 75\% was measured as the laser frequency was scanned, with output powers up to 110\,mW. The device was locked stably over an hour using a PDH scheme producing 28\,mW of second harmonic power (conversion efficiency greater than 50\%) via a laser frequency feedback scheme. The results presented here showcase the suitability of titanium indiffused waveguide resonators for high power, low loss applications.

\section*{Funding}
The authors acknowledge funding from DFG (Deutsche Forschungsgemeinschaft) via the Gottfried Wilhelm Leibniz-Preis.


\begin{thebibliography}{10}

\bibitem{Wooten00.JQE}
E.~L. Wooten, K.~M.~K. nad A.~Yi-Yan, E.~J. Murphy, D.~A. Lafaw, P.~F.
  Hallemeier, D.~Maack, D.~V. Attanasio, D.~J. Fritz, G.~J. McBrien, and D.~E.
  Bossi, \newblock{A review of lithium niobate modulators for fiber-optic
  communications systems,} {IEEE} J. Sel. Topics Quantum Electron. \textbf{6},
  69 (2000).

\bibitem{Arbore97.OL}
M.~A. Arbore and M.~M. Fejer, \newblock{Singly resonant optical parametric
  oscillation in periodically poled lithium niobate waveguides,} Opt. Lett.
  \textbf{22}, 151 (1997).

\bibitem{Anderson95.OL}
M.~E. Anderson, M.~Beck, M.~G. Raymer, and J.~D. Bierlein, \newblock{Quadrature
  squeezing with ultrashort pulses in nonlinear-optical waveguides,} Opt. Lett.
  \textbf{20}, 620 (1995).

\bibitem{Stefszky17.PRA}
M.~Stefszky, R.~Ricken, C.~Eigner, V.~Quiring, H.~Herrmann, and C.~Silberhorn,
  \newblock{Waveguide cavity resonator as a source of optical squeezing,} Phys.
  Rev. Applied \textbf{7}, 044026 (2017).

\bibitem{Ruetz16.APB}
H.~R\"{u}tz, K.-H. Luo, H.~Suche, and C.~Silberhorn, \newblock{Towards a quantum
  interface between telecommunication and {UV} wavelengths: {D}esign and
  classical performance,} Appl. Phys. B \textbf{122}, 13 (2016).

\bibitem{Iwai03.APL}
M.~Iwai, T.~Yoshino, S.~Yamaguchi, M.~Imaeda, N.~Pavel, I.~Shoji, and T.~Taira,
  \newblock{High-power blue generation from a periodically poled {MgO:LiNbO$_3$}
  ridge-type waveguide by freqeuncy doubling of a diode end-pumped
  {Nd:Y$_3$Al$_5$O$_12$} laser,} App. Phys. Lett. \textbf{83}, 3659 (2003).

\bibitem{Vandevender.JOMO}
A.~P. Vandevender and P.~Kwiat, \newblock{High efficiency single photon
  detection via frequency up-conversion,} J. Mod. Opt. \textbf{51}, 1433
  (2004).

\bibitem{Vollmer14.PRL}
C.~E. Vollmer, C.~Braune, A.~Samblowski, T.~Eberle, V.~H\"{a}ndchen,
  J.~Fiur\'{a}\v{s}ek, and R.~Schnabel, \newblock{Quantum up-conversion of
  squeezed vacuum states from 1550 to 532 nm,} Phys. Rev. Lett. \textbf{112},
  073602 (2014).

\bibitem{Ulvila13.OL}
V.~Ulvila, C.~R. Phillips, L.~Halonen, and M.~Vainio, \newblock{Frequency comb
  generation by a continuous-wave-pumped optical parametric oscillator based on
  cascading quadratic nonlinearities,} Opt. Lett. \textbf{38}, 4281 (2013).

\bibitem{Ricciardi15.PRA}
I.~Ricciardi, S.~Mosca, M.~Parisi, P.~Maddaloni, L.~Santamaria, P.~D. Natale,
  and M.~D. Rosa, \newblock{Frequency comb generation in quadratic nonlinear
  media,} Phys. Rev. A \textbf{91}, 063839 (2015).

\bibitem{Leo16.PRL}
F.~Leo, T.~Hansson, I.~Ricciardi, M.~D. Rosa, S.~Coen, S.~Wabnitz, and
  M.~Erkintalo, \newblock{Walk-off-induced modulational instability, temporal
  pattern formation, and frequency comb generation in cavity-enhanced
  second-harmonic generation,} Phys. Rev. Lett. \textbf{116}, 033901 (2016).

\bibitem{Sun12.OL}
J.~Sun and C.~Xu, \newblock{466 m{W} green light generation using annealed
  proton-exchanged periodically poled {MgO:LiNbO$_3$} ridge waveguides,} Opt.
  Lett. \textbf{37}, 2028 (2012).

\bibitem{Mizuuchi03.OL}
K.~Mizuuchi, T.~Sugita, K.~Yamamoto, T.~Kawaguchi, T.~Yoshino, and M.~Imaeda,
  \newblock{Efficient 340-nm light generation by a ridge-type waveguide in a
  first-order periodically poled {MgO:LiNbO$_3$},} Opt. Lett. \textbf{28}, 1344
  (2003).

\bibitem{Sakai07.OL}
K.~Sakai, Y.~Koyata, and Y.~Hirano, \newblock{Blue light generation in a ridge
  waveguide {MgO:LiNbO$_3$} crystal pumped by a fiber {B}ragg grating
  stabilized laser diode,} Opt. Lett. \textbf{32}, 2342 (2007).

\bibitem{Sakai06.OL}
K.~Sakai, Y.~Koyata, and Y.~Hirano, \newblock{Planar-waveguide
  quasi-phase-matched second-harmonic-generation device in {Y}-cut {MgO}-doped
  {LiNbO$_3$},} Opt. Lett. \textbf{31}, 3134 (2006).

\bibitem{Luo15.NJP}
K.-H. Luo, H.~Herrmann, S.~Krapick, B.~Brecht, R.~Ricken, V.~Quiring, H.~Suche,
  W.~Sohler, and C.~Silberhorn, \newblock{Direct generation of
  single-longitudinal-mode narrowband photon pairs,} New J. Phys. \textbf{17},
  073039 (2015).

\bibitem{Regener88.JOSAB}
R.~Regener and W.~Sohler, \newblock{Efficient second-harmonic generation in
  {Ti:LiNbO}$_3$ waveguide resonators,} J. Opt. Soc. Am. B \textbf{5}, 267
  (1998).

\bibitem{Pernice12.APL}
W.~H.~P. Pernice, C.~Xiong, C.~Schuck, and H.~X. Tang, \newblock{Second harmonic
  generation in phase matched alumninum nitride waveguides and micro-ring
  resonators,} Appl. Phys. Lett.  (2012).

\bibitem{Scaccabarozzi06.OL}
L.~Scaccabarozzi, M.~M. Fejer, Y.~Huo, S.~Fan, X.~Yu, and S.~Harris,
  \newblock{Enhanced second-harmonic generation in {AlGaAs/Al$_x$O$_y$} tightly
  confining waveguides and resonant cavities,} Optics Lett.  (2006).

\bibitem{Fujimura96.JQE}
M.~Fujimura, M.~Sudoh, K.~Kintaka, T.~Suhara, and H.~Nishihara,
  \newblock{Resonant waveguide quasi-phase-matched {SHG} devices with
  electrooptic phase-modulator for tuning,} IEEE J. Sel. Top. Quantum Electron.
  \textbf{2}, 396 (1996).

\bibitem{Gui09.OE}
L.~Gui, H.~Hu, and W.~S. M.~Garcia-Granda, \newblock{Local periodic poling of
  ridges and ridge waveguides on {X}- and {Y}-{C}ut {LiNbO$_3$} and its
  application for second harmonic generation,} Nat. Phys. \textbf{5}, 541
  (2008).

\bibitem{Nava13.APL}
S.~Nava, P.~Minzioni, I.~Cristiani, N.~Argiolas, M.~Bazzan, M.~V. Ciampolillo,
  G.~Pozza, C.~Sada, and V.~Degiorgio, \newblock{Photorefractive effect at 775nm
  in doped lithium niobate crystals,} Appl. Phys. Lett. \textbf{103}, 031904
  (2013).

\bibitem{Pal15.APB}
S.~Pal, B.~K. Das, and W.~Sohler, \newblock{Photorefractive damage resistance in
  {Ti:PPLN} waveguides with ridge geometry,} Appl. Phys. B \textbf{120}, 737
  (2015).

\bibitem{Becker85.APL}
R.~A. Becker and R.~C. Williamson, \newblock{Photorefractive effects in
  {LiNbO}$_3$ channel waveguides: {M}odel and experimental verification,} Appl.
  Phys. Lett. \textbf{10}, 1024 (1985).

\bibitem{Kondo.AO}
Y.~Kondo, S.~Miyaguchi, A.~Onoe, and Y.~Fujii, \newblock{Quantitatively measured
  photorefractive sensitivity of proton-exchanged lithium niobate,
  proton-exchanged magnesium oxide-doped lithium niobate, and ion-exchanged
  potassium titanyl phosphate waveguides,} Appl. Opt. \textbf{33}, 3348 (1994).

\bibitem{Carrascosa08.OE}
M.~Carrascosa, J.~Villaeroel, J.~Carnicero,
  A.~Garcia-Caba$\tilde{\textrm{n}}$es, and J.~M. Cabrera,
  \newblock{Understanding light intensity thresholds for catastrophic optical
  damage in {LiNbO}$_3$,} Opt. Exp. \textbf{16}, 115 (2008).

\bibitem{Black01.AJP}
E.~D. Black, \newblock{An introduction to pound-drever-hall laser frequency
  stabilization,} Am. J. Phys. \textbf{69}, 79 (2001).

\bibitem{Stefszky12.CQG}
M.~S. Stefszky, C.~M. Mow-Lowry, S.~S.~Y. Chua, D.~A. Shaddock, B.~C. Buchler,
  H.~Vahlbruch, A.~Khalaidovski, R.~Schnabel, P.~K. Lam, and D.~E. McClelland,
  \newblock{Balanced homodyne detection of optical quantum states at audio-band
  frequenices and below,} Class. Quantum Grav.  (2012).

\bibitem{Mehmet11.OE}
M.~Mehmet, S.~Ast, T.~Eberle, S.~Steinlechner, H.~Vahlbruch, and R.~Schnabel,
  \newblock{Squeezed light at 1550nm with a quantum noise reduction of
  12.3$\,$d{B},} Opt. Exp. \textbf{19}, 25763 (2011).

\bibitem{Carrascosa93.JAP}
M.~Carrascosa and L.~Arizmendi, \newblock{High-temperature photorefractive
  effects in {LiNbO$_{3}$:Fe}}, J. Appl. Phys. \textbf{73}, 2709, (1993).

\bibitem{Imeshev98.OL}
G.~Imeshev, M.~Proctor, and M.~M. Fejer, \newblock{Phase correction in
  double-pass quasi-phase-matched second-harmonic generation with a wedged
  crystal,} Opt. Lett. \textbf{23}, 165 (1998).

\bibitem{Siegman}
A.~E. Siegman, \emph{Lasers} (University Science Books, Mill Valley,
  California, 1986).

\bibitem{Shoji97.JOSAB}
I.~Shoji, T.~Kondo, A.~Kitamoto, M.~Shirane, and R.~Ito, \newblock{Absolute
  scale of seond-order nonlinear-optical coefficients,} J. Opt. Soc. Am. B
  \textbf{14}, 2268 (1997).

\bibitem{Fujimura96.JLT}
M.~Fujimura, T.~Suhara, and H.~Nishihara, \newblock{Theoretical analysis of
  resonant waveguide optical second harmonic generation devices,} J. Lightwave
  Technol. \textbf{14}, 1899 (1996).

\bibitem{Fujiwara89.APL}
T.~Fujiwara, S.~Sato, and H.~Mori, \newblock{Wavelength dependence of
  photorefractive effect in {T}i-indiffused {LiNbO$_{3}$} waveguides,} Appl.
  Phys. Lett. \textbf{54}, 975 (1989).

\end{thebibliography}
\end{document}